\documentclass[a4paper,twocolumn,english,prl, superscriptaddress, preprintnumbers, showpacs, nofootinbib]{revtex4}
\usepackage[T1]{fontenc}
\usepackage[latin1]{inputenc}
\usepackage{float}
\usepackage{amsmath}
\usepackage{graphicx}
\usepackage{amssymb}

\makeatletter
\usepackage{amsfonts}
\usepackage{latexsym}
\usepackage[T1]{fontenc}
\usepackage{ae,aecompl}

\usepackage{babel}
\makeatother
\begin{document}

\preprint{EPFL-ITS-37.2005}

\title{Global universe anisotropy probed by the alignment of structures
in the cosmic microwave background}

\author{Y. Wiaux}

\email{yves.wiaux@epfl.ch}

\affiliation{Signal Processing Institute, Ecole Polytechnique Fédérale de Lausanne
(EPFL), CH-1015 Lausanne, Switzerland}

\author{P. Vielva}

\email{vielva@ifca.unican.es}

\affiliation{Instituto de F\'isica de Cantabria (CSIC-UC), E-39005 Santander,
Spain}

\author{E. Mart\'inez-Gonz\'alez}

\affiliation{Instituto de F\'isica de Cantabria (CSIC-UC), E-39005 Santander,
Spain}

\author{P. Vandergheynst}

\affiliation{Signal Processing Institute, Ecole Polytechnique Fédérale de Lausanne
(EPFL), CH-1015 Lausanne, Switzerland}

\begin{abstract}
We question the global universe isotropy by probing the alignment
of local structures in the cosmic microwave background (CMB) radiation.
The original method proposed relies on a steerable wavelet decomposition
of the CMB signal on the sphere. The analysis of the first-year WMAP
data identifies a mean preferred plane with a normal direction close
to the CMB dipole axis, and a mean preferred direction in this plane,
very close to the ecliptic poles axis. Previous statistical anisotropy
results are thereby synthesized, but further analyses are still required
to establish their origin.
\end{abstract}


\maketitle

\section{Introduction}

Last years' experiments in cosmology have resulted in the definition
of a consistent picture of the structure and evolution of the universe.
The recent data of the cosmic microwave background (CMB) radiation,
together with other cosmological observations, have allowed us to
determine precise values for the main cosmological parameters \cite{CMBspergel}.
However, the corresponding concordance cosmological model is based
on strong hypotheses which need to be questioned. They extend from
the nature of the gravitational interaction underpinning the cosmological
evolution of the universe, to the physics governing the early inflationary
era, or also to the cosmological principle for the global homogeneity
and isotropy of the universe.

This letter defines an original method to test the global universe
isotropy through the analysis of the CMB data. The observed CMB anisotropies
on the celestial sphere can be interpreted as a realization of a statistical
process originating in the inflationary era. The cosmological principle
implies the isotropy of the corresponding statistical properties.
This statistical isotropy has been challenged through multiple methods
applied to the first year full-sky CMB data of the Wilkinson Microwave
Anisotropy Probe (WMAP) experiment. The detections quoted in the following
suggest statistical anisotropy with confidence levels higher than
$99$ percent. Analyses based on N-point correlation functions \cite{SAeriksen1,SAeriksen2},
local curvature \cite{SAhansen2}, local power spectra \cite{SAhansen1,SAhansen3,SAhansen4,SAdonoghue},
and bispectra \cite{NGland4}, suggest a north-south asymmetry maximized
in a coordinate system with the north pole at $(\theta,\varphi)=(80^{\circ},57^{\circ})$
in Galactic co-latitude $\theta$ and longitude $\varphi$, close
to the north ecliptic pole lying at $(\theta,\varphi)=(60^{\circ},96^{\circ})$.
Analyses of multipole vectors, angular momentum dispersion, as well
as azimuthal phases correlations find an anomalous alignment between
the low $l$ multipoles of the CMB, suggesting a preferred direction
around $(\theta,\varphi)=(30^{\circ},260^{\circ})$, near the ecliptic
plane and close to the axis of the dipole lying at $(\theta,\varphi)=(42^{\circ},264^{\circ})$
\cite{SAcopi,SAkatz,SAschwarz,NGland1,NGland3,SAoliveiracosta,NGland2,SAbielewicz}.
Galactic north-south asymmetries are also found in the analysis of
the kurtosis and the area of the wavelet coefficients of the CMB data.
These are mainly due to a very cold spot in the southern hemisphere
\cite{WNGvielva,WNGcruz}. First results with the angular pair separation
method, which probes the statistical isotropy both in real and multipole
space, also seem to support those results \cite{SAbernui}. On the
contrary, bipolar power spectra analyses are consistent with no violation
of the statistical isotropy of the universe \cite{SAhajian3,SAhajian1,SAhajian2}.
Finally, theoretical models for an anisotropic universe are being
studied to account for the observed effects \cite{SAJaffe}.

Our alternative method probes the statistical isotropy of the CMB
by the analysis of the alignment of structures in the signal. Preferred
directions in the universe are defined as the directions towards which
local features of the CMB are mostly oriented. The level of preference
of each direction may be established from simulations, and is represented
as a signal on the sphere. The approach is therefore powerful as it
also \emph{a priori} allows the study of the corresponding angular
power spectrum in order to probe the multipole distribution of the
anisotropy. The analysis defined relies on a steerable wavelet decomposition
of the first year full-sky WMAP data.

\section{Data and simulations}

The experimental temperature map used for the analysis is obtained
from the first year WMAP data following the procedure originally proposed
in \cite{NGkomatsu2} for gaussianity tests. First, a best estimation
of diffuse galactic foregrounds is removed from each of the eight
frequency maps of the WMAP receivers at the $Q$, $V$, and $W$ bands,
as prescribed by the specific template fits method in \cite{CMBbennet2}.
The eight foreground cleaned maps are then combined through a noise-weighted
linear combination in order to enhance the CMB signal-to-noise ratio.
The so-called Kp0 mask of \cite{CMBbennet2} is applied to account
for the remaining strong foreground contamination by diffuse galactic
emissions around the galactic plane and by bright point sources. Finally,
the residual monopole and dipole are subtracted outside the mask.
All the WMAP data are released in the HEALPix pixelization on the
sphere \cite{SASgorski}, at a resolution identified by the parameter
$N_{side}=512$, corresponding to maps with several million pixels
with a spatial resolution of $6.9$ arcminutes. The temperature map
obtained from the above pre-processing is downgraded to a lower resolution
with $N_{side}=32$ for the specific purpose of our analysis. This
provides a map with $N_{pix}=12288$ pixels with a spatial resolution
of $1.8$ degrees.

Ten thousand temperature map simulations at the resolution $N_{side}=32$
are used in order to define confidence levels for the analysis results.
These simulations specifically assume the gaussianity and statistical
isotropy of the CMB. They are generated following the scheme also
proposed in \cite{NGkomatsu2}. A gaussian CMB simulation is obtained
in spherical harmonics space from the angular power spectrum determined
by the cosmological parameters of the WMAP best-fit cosmological model.
The observation at each receiver is simulated by convolving that map
with the corresponding WMAP beam window function. After transforming
each map to pixel space, a gaussian noise realization is also added
with the proper dispersion per pixel. Following the same procedure
as for the data, each one of the simulated frequency maps of the eight
receivers at the $Q$, $V$, and $W$ bands are finally combined through
the noise-weighted linear combination which enhances the CMB signal-to-noise
ratio, the Kp0 mask is applied, and the monopole and dipole are subtracted.

\section{Wavelet filtering}

We recall the recently introduced steerable wavelet decomposition
of a signal on the sphere upon which we base our analysis.

Let the function $F(\omega)$ describe a signal on the sphere, with
the point $\omega=(\theta,\varphi)$ identified by its co-latitude
$\theta\in[0,\pi]$ and longitude $\varphi\in[0,2\pi)$. A wavelet
on the sphere is a localized function $\Psi$ that can be dilated
at any scale $a\in\mathbb{R}_{+}^{*}$ associated with a given angular
opening on the sphere, rotated on itself by any angle $\chi\in[0,2\pi)$,
and finally translated at any position $\omega_{0}=(\theta_{0},\varphi_{0})$.
The wavelet filtering of $F$ by $\Psi$ results from the scalar product
(correlation) of the signal with all the dilated, rotated, and translated
versions of the wavelet: $\Psi_{\omega_{0},\chi,a}$. Hence, the wavelet
coefficients $W_{\Psi}^{F}(\omega_{0},\chi,a)=\langle\Psi_{\omega_{0},\chi,a}|F\rangle$
characterize the signal locally at each scale $a$, in each orientation
$\chi$, and at each point $\omega_{0}$ \cite{WSantoine1,SASwiaux1}.
Individualizing the properties of the signal independently at each
scale is a major advantage of the wavelet scale-space filtering as
physical phenomena may generally be scale dependent. Determining the
local orientations of the signal structure at each scale also provides
essential information for our particular analysis.

\begin{figure}[H]
\begin{center}\includegraphics[%
  height=3cm,
  keepaspectratio,
  angle=-90]{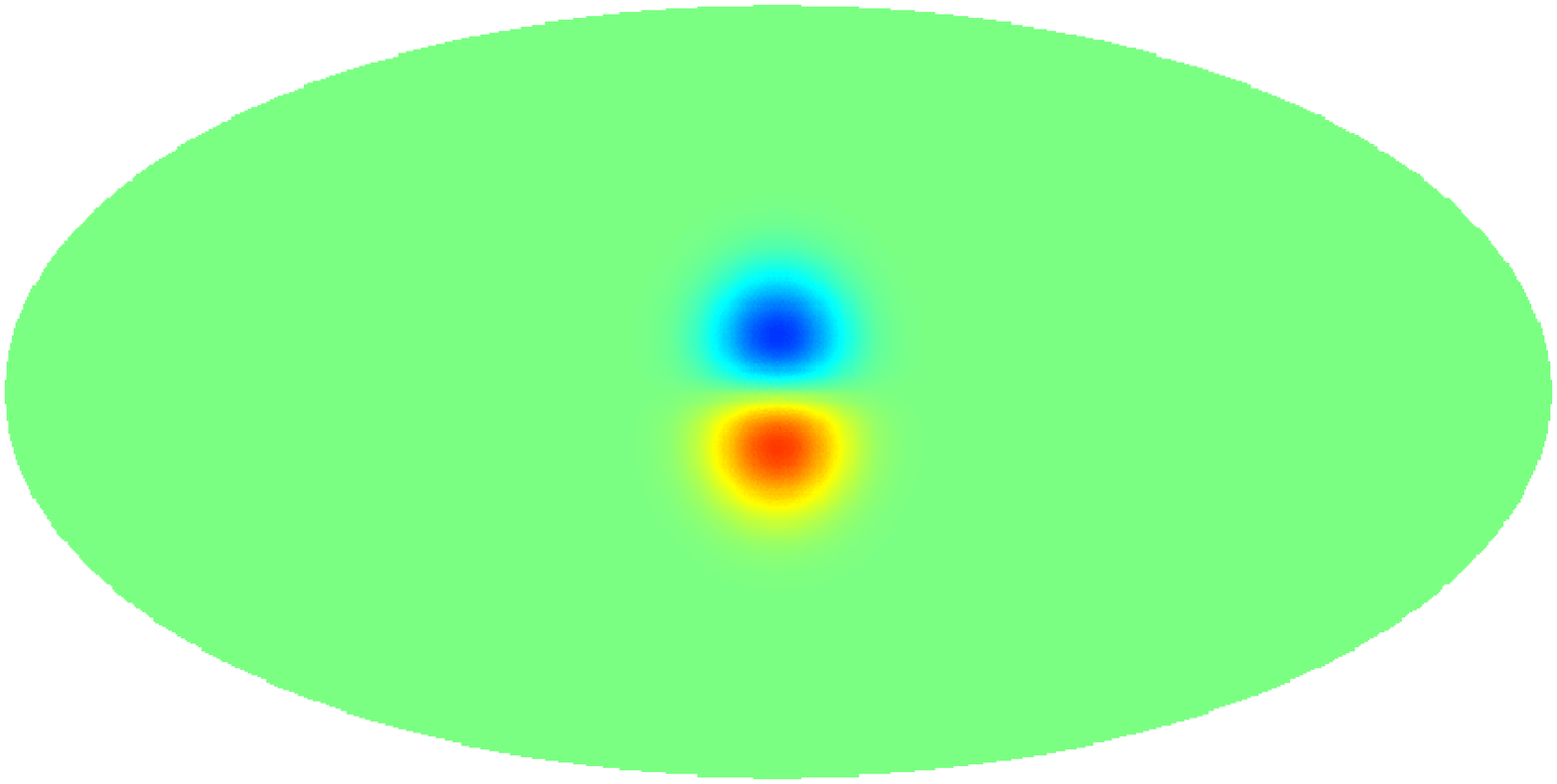}\hspace{1cm}\includegraphics[%
  height=3cm,
  keepaspectratio,
  angle=-90]{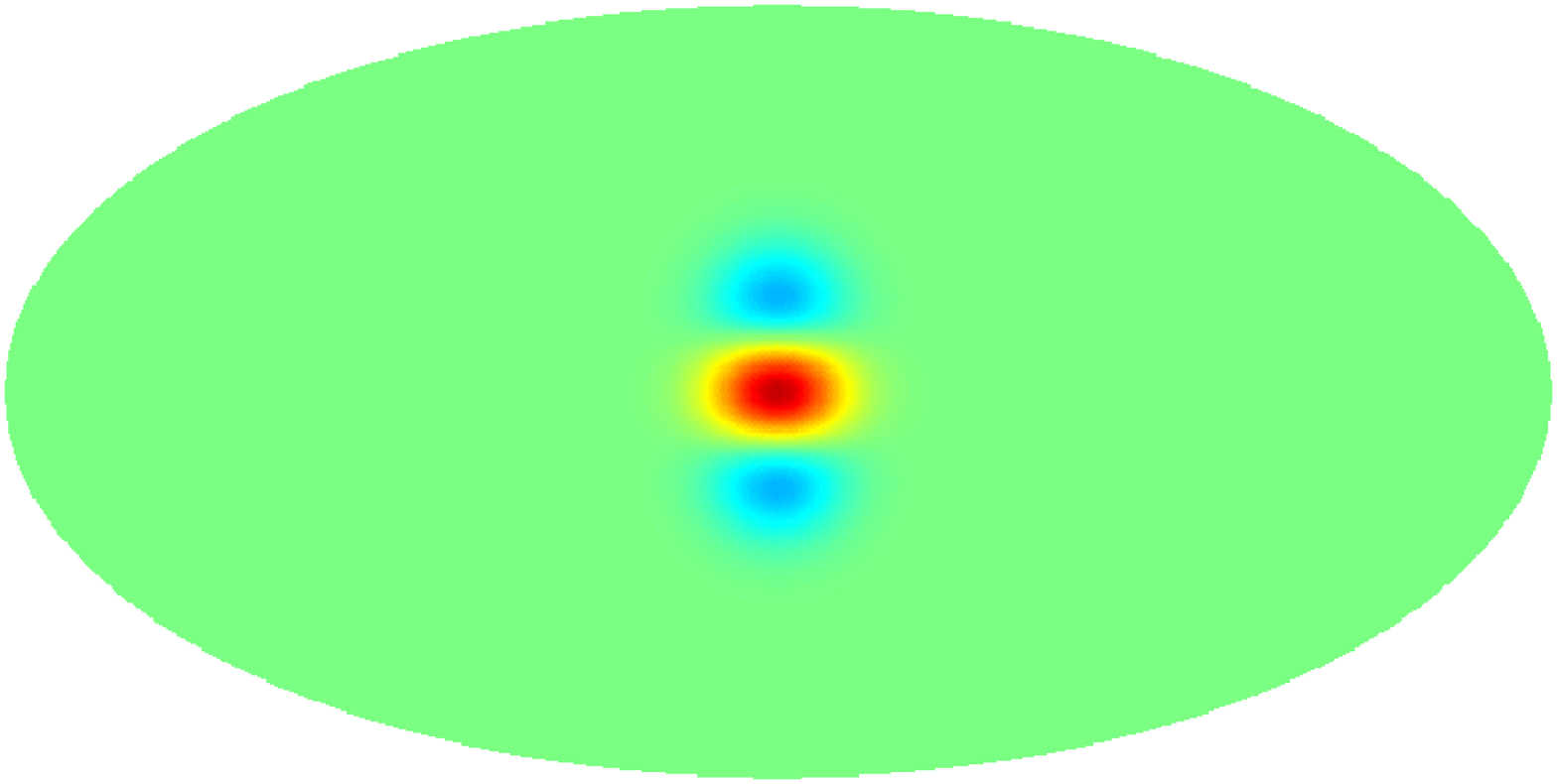}\end{center}

\caption{\label{cap:gaussianderivatives}Mollweide projections of the first
(left) and second (right) gaussian derivative wavelets at position
$\omega_{0}=(\pi/2,0)$, orientation $\chi=0$, and scale $a=0.19$.
Red and blue regions respectively correspond to positive and negative
values.}
\end{figure}

Notice that the computation of all wavelet coefficients for a large
number of scales, and with the same high resolution in local orientation
and position is a numerically complex calculation. A corresponding
fast directional correlation algorithm on the sphere was recently
defined which strongly relies on the use of so-called steerable wavelet
filters \cite{SASwiaux3}. Our analysis of the WMAP data is performed
independently with the first and second gaussian derivatives. These
steerable filters, represented in Mollweide projection in figure \ref{cap:gaussianderivatives},
are analytically defined in \cite{SASwiaux1,SASwiaux3}. %

\section{Analysis method}

Here, we consider the analysis leading to the identification of preferred
directions from the first year full-sky WMAP data at the HEALPix resolution
$N_{side}=32$. The complementary study of the corresponding angular
power spectrum is postponed to a future work.

For a given scale $a$, the angular size of our wavelets on the sphere
is defined as twice the dispersion of the corresponding gaussian.
Twelve scales are selected corresponding to angular sizes of the first
and second gaussian derivatives lying between $5$ and $30$ degrees.
At each scale, an extended exclusion mask $M_{a}$ is defined in order
to avoid considering pixels for which the correlation between the
WMAP signal and the wavelet is contaminated by the Kp0 mask. The pixels
added to the initial mask are those for which the wavelet coefficients
of a constant signal in the region of the mask are non-zero, up to
a given threshold.

For each pixel $\omega_{0}$ of the signal outside the extended exclusion
mask, the direction $\chi_{0}(\omega_{0})$ for which the wavelet
coefficient is maximum in absolute value is selected, and the corresponding
absolute value is retained. This selects the local wavelet orientation
which best matches the orientation of the local structure of the signal
at each point. The great circle is defined which passes through that
point and admits the corresponding local orientation as a tangent.
The directions in the sky lying on that great circle are considered
to be highlighted by the local structure identified, and are weighted
by the absolute value of the wavelet coefficient $\vert W_{\Psi}^{F}(\omega_{0},\chi_{0}(\omega_{0}),a)\vert$. 

Consider each direction $\omega$ on the celestial sphere in the set
$S$ of the $N_{pix}=12288$ pixels at $N_{side}=32$. The total weight
$D_{a}(\omega)$ at scale $a$ is the sum of the $N_{cros}(\omega)$
weights originating from all pixels $\omega_{0}^{(c)}$ in the original
signal, with $1\leq c\leq N_{cros}(\omega)$, for which the great
circle defined crosses the direction considered: \begin{equation}
D_{a}\left(\omega\right)=\frac{1}{A}\sum_{c=1}^{N_{cros}(\omega)}\vert W_{\Psi}^{F}\left(\omega_{0}^{(c)},\chi_{0}(\omega_{0}^{(c)}),a\right)\vert.\label{2}\end{equation}
 The factor $A=LN_{pix}^{-1}\sum_{\omega_{0}\notin M_{a}}\vert W_{\Psi}^{F}(\omega_{0},\chi_{0}(\omega_{0}),a)\vert$
of normalization defines a mean total weight in each direction equal
to unity for isotropic CMB simulations without mask: $N_{pix}^{-1}\sum_{\omega\in S}D_{a}(\omega)=1$.
The quantity $L=4N_{side}$ stands for the number of points on a great
circle on a HEALPix grid. Preferred directions in the universe are
therefore identified as the most weighted ones. Notice that the procedure
obviously assigns identical total weights to opposite directions.
In other words, our directions are headless vectors.

\section{Results}

In this section, the results of the analysis of the first year WMAP
data proposed above are exposed. The filtering of the CMB data by
the first gaussian derivative does not lead to any significant detection.
This suggests that there is no clear alignment of the local CMB structures
with the morphology captured by that specific filter. The corresponding
analysis with the second gaussian derivative leads to a strong detection,
detailed in the following.

\begin{figure}[H]
\begin{center}\includegraphics[%
  height=8cm,
  keepaspectratio,
  angle=-90]{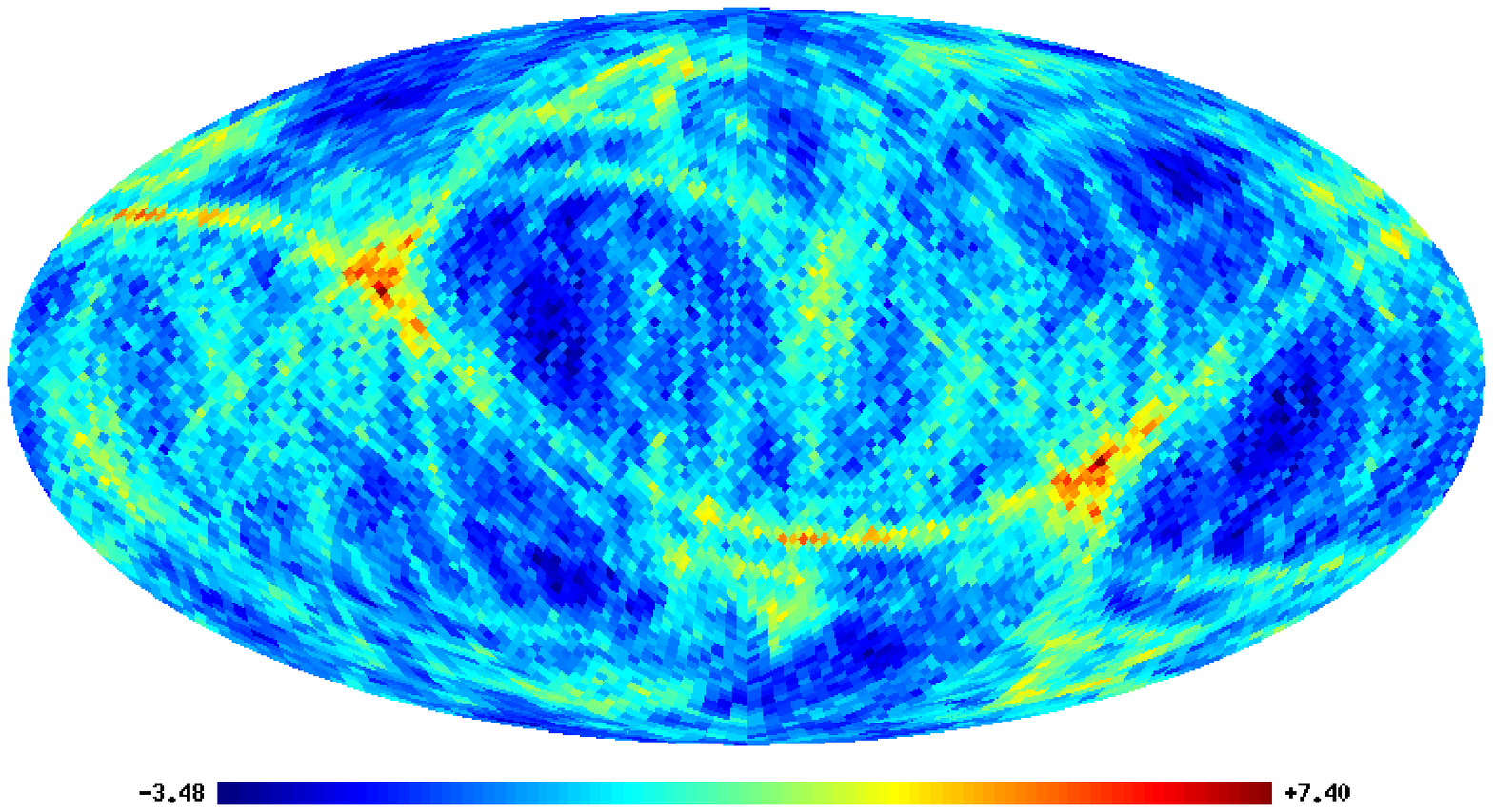}\\
\includegraphics[%
  height=8cm,
  keepaspectratio,
  angle=-90]{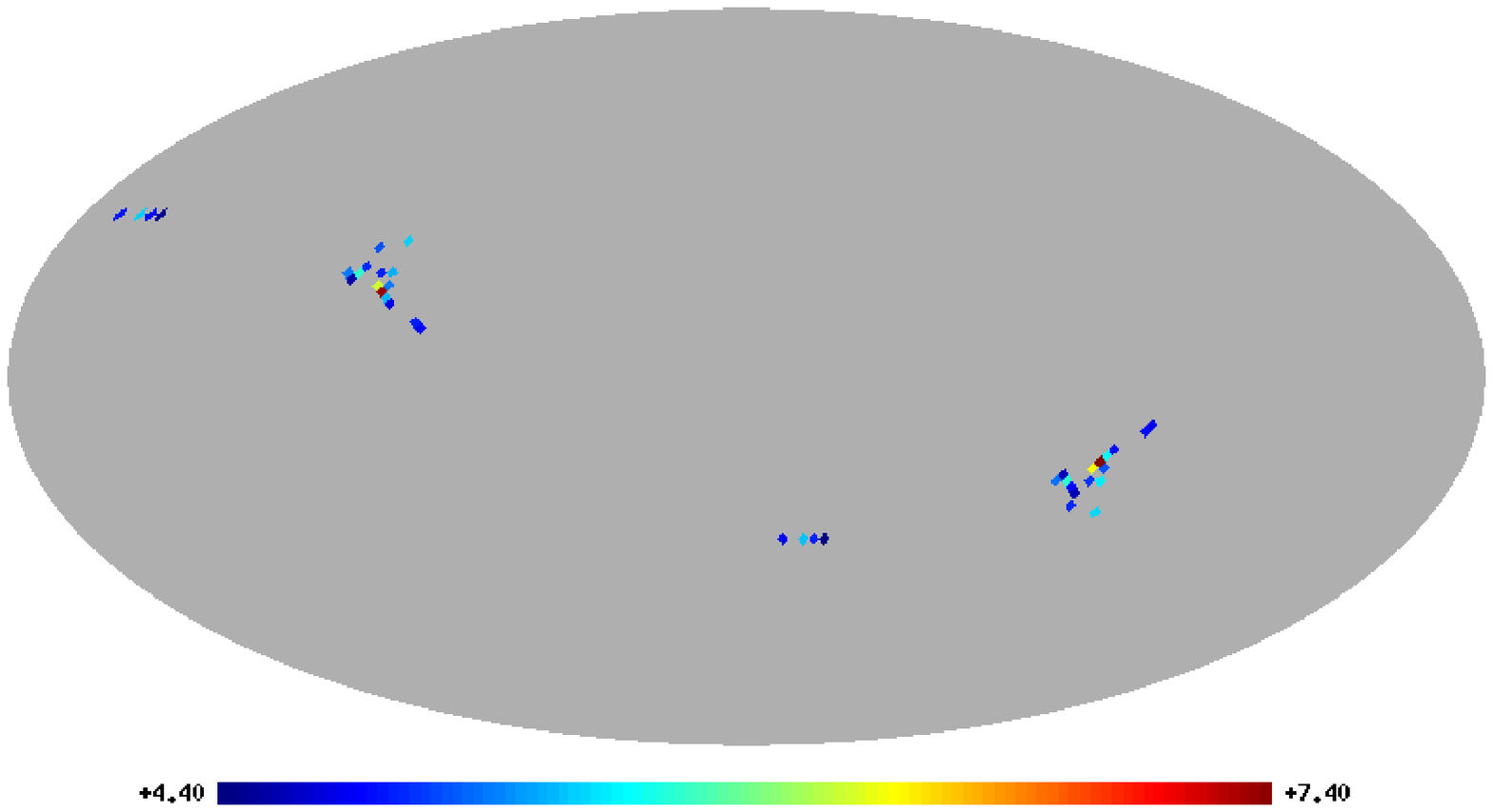}\end{center}

\caption{\label{cap:totalweights-anomalies}Mollweide projection of the galactic
coordinates maps of total weights (top) and directions anomalous at
$99.99$ percent (bottom) at the scale $a_{3}$ of the second gaussian
derivative wavelet, and in $\sigma_{a_{3}}$ units.}
\end{figure}

At each scale $a$ and at each pixel $\omega$, the total weight obtained
can be quantified by the number of standard deviations $\sigma_{a}(\omega)$
through which it deviates from the mean $\mu_{a}(\omega)$, as estimated
from the simulations. Preferred directions, or positive total weights,
correspond to $D_{a}(\omega)>\mu_{a}(\omega)$. Non-preferred directions,
or negative total weights, are distinguished as $D_{a}(\omega)<\mu_{a}(\omega)$.
We consider in particular the scale $a_{3}$ associated with an angular
size of the filter of $8.3$ degrees on the sky. The corresponding
map of total weights clearly depicts the distribution of anisotropy
resulting from our analysis (figure \ref{cap:totalweights-anomalies},
top panel).

A rare detection is observed at that specific scale. It identifies
$20$ directions (pairs of opposite points), qualified as anomalous
at $99.99$ percent, with an associated positive total weight higher
than in any of the ten thousand simulations considered. The $20$
anomalous directions obtained at scale $a_{3}$, with positive total
weights lying in the interval $[4.4,7.4]$ in $\sigma_{a_{3}}(\omega)$
units, are mainly concentrated in two clusters around the ecliptic
poles (figure \ref{cap:totalweights-anomalies}, bottom panel). The
most prominent of these concentrations comprises $16$ of the $20$
anomalous directions. Its mean position is identified by a northern
end at $(\theta,\varphi)=(71^{\circ},91^{\circ})$ in galactic co-latitude
$\theta$ and longitude $\varphi$, when each direction is weighted
by its corresponding total weight. This defines a mean preferred direction
in the sky with a northern end very close to the north ecliptic pole
at $(\theta,\varphi)=(60^{\circ},96^{\circ})$. The presence of the
two elongated clusters also suggests that the $20$ anomalous directions
lie on a great circle, defining a preferred plane in the sky. The
normal direction to the mean preferred plane, given by the cross-product
of the mean directions of each of the two clusters, has a northern
end position at $(\theta,\varphi)=(34^{\circ},331^{\circ})$. Alternatively,
the great circle which best fits the $20$ anomalous directions identifies
exactly the same plane. This normal direction lies close to the northern
end of the CMB dipole axis at $(\theta,\varphi)=(42^{\circ},264^{\circ})$.

Let us emphasize that $11$ directions anomalous at $99.99$ percent
are also detected at the neighbour scale $a_{4}$ corresponding to
an angular size of the filter of $10$ degrees. They belong to two
clusters with very similar mean positions to those identified at scale
$a_{3}$, and therefore single out the same structure of anisotropy.%

\section{Discussion and conclusion}

The present original analysis of structures alignment in the CMB from
the first year WMAP data clearly identifies a mean preferred plane
in the universe with a normal direction close to the CMB dipole axis,
and a mean preferred direction in this plane, very close to the ecliptic
poles axis. This result is based on the observation of $20$ directions
anomalous at $99.99$ percent. Our original method thus singles out
the same directions as those highlighted by previous statistical isotropy
studies. The wavelet approach also identifies the angular size of
the anomalously aligned structures around $8.3$ degrees on the celestial
sphere, corresponding to a multipole range roughly between $l=11$
and $l=27$.

This new insight into the anisotropy structure might help us to understand
its still unclear origin. First, the angular size identified for the
aligned structures is compatible with the size of primary CMB anisotropies
due to topological defects such as texture fields \cite{CMBturok}
or secondary anisotropies due to the Rees-Sciama effect \cite{CMBmartinez}.
Alignment mechanisms \cite{SAgordon,SArakic} were recently proposed
which might be generalized to such structures. Second, the identified
angular size is also compatible with the one above which diffuse galactic
emissions dominate the CMB emission. Corresponding residual foregrounds
can be probed by a direct analysis of the WMAP $Q$, $V$, and $W$
bands frequency maps. Finally, the coincidence of the preferred directions
detected with the ecliptic poles and dipole axes naturally suggests
possible unknown systematic effects \cite{SAgordon,SAfreeman}. In
that respect, notice that the angular size of the mesh of the WMAP
scan pattern defined by the combination of the spin and precession
of the satellite is, again, of the order of several degrees \cite{CMBbennet1}.

In conclusion, our analysis provides a first synthesis of previous
statistical anisotropy results. Further analyses of the complete total
weights distributions at various wavelet scales and of the corresponding
angular power spectra can allow a deeper probe of the anisotropy structure.
But the origin of the present detection already needs to be thoroughly
investigated. Various hypotheses can be suggested in terms of cosmological
or foreground structures, or systematics. But nothing at present allows
us to discard the possibility of a global universe anisotropy, simple
violation of the cosmological principle hypothesis.

\begin{acknowledgments}
The authors acknowledge the use of the LAMBDA archive, and of the
HEALPix and CMBFAST softwares. Y. W. was supported by the Swiss and
Belgian National Science Foundations. P. V. and E. M.-G. were supported
by the Spanish MEC project ESP2004-07067-C03-01.
\end{acknowledgments}

\end{document}